\documentclass{aa}

\usepackage{graphicx}
\usepackage{txfonts}
\usepackage{natbib}
\usepackage[
    dvips, 
    latex2html,
    colorlinks=true,
    urlcolor=blue,
    bookmarks,
    bookmarksnumbered,
    pdftitle={A great document!},
    pdfauthor={JaZ},
    pdfsubject={Producing high-quality pdf documents with LaTeX},
    pdfcreator={JZ},
    pdfproducer={JC},
    pdfkeywords={LaTeX, pdf, bookmarks, hyperlinks},
    pdfcenterwindow=true
    ]{hyperref}

\def\tr{\textrm}
\newcommand{\tty}[1]{\textrm{\tiny #1}}
\newcommand{\unit}[2]{\tr{#1}^{#2}}

\def\INTEGRAL{\emph{INTEGRAL}}
\def\XMM{\emph{XMM-Newton}}

\def\IGR{\object{IGR~J17252$-$3616}}
\def\Msol{$M_{\odot}$}
\def\cps{counts$\,\unit{s}{-1}$}

\begin{document}
\title{IGR~J17252$-$3616: an accreting pulsar observed by \INTEGRAL\ and \XMM}

\titlerunning{IGR~J17252$-$3616 observed by \INTEGRAL\ and \XMM}


\author{J.A. Zurita Heras\inst{1}\fnmsep\inst{2}
   \and G. De Cesare\inst{3}\fnmsep\inst{4}\fnmsep\inst{5}
   \and R. Walter \inst{1}\fnmsep\inst{2}
   \and A. Bodaghee \inst{1}\fnmsep\inst{2}
   \and G. B\'elanger \inst{8}
   \and T.J.-L. Courvoisier \inst{1}\fnmsep\inst{2}
   \and S.E. Shaw \inst{6}\fnmsep\inst{1}
   \and J.B. Stephen \inst{7}
       }
\authorrunning{J.A. Zurita Heras \emph{et al.}}

\offprints{J.A. Zurita Heras\\ \email{Juan.Zurita@obs.unige.ch}}

\institute{INTEGRAL Science Data Centre, ch. d'Ecogia 16, 1290 Versoix, Switzerland
      \and Observatoire de Gen\`eve, ch. des Maillettes 51, 1290 Sauverny, Switzerland
      \and IASF-INAF, Via Fosso del Cavaliere 100, 00133 Roma, Italy
      \and Dipartimento di Astronomia, Universita degli Studi di Bologna, Via Ranzani 1, I40127 Bologna, Italy
      \and Centre d'Etude Spatiale des Rayonnements, CNRS/UPS, B.P. 4346, 31028 Toulouse Cedex 4, France 
      \and School of Physics and Astronomy, University of Southampton, Highfield, SO17 1BJ, UK
      \and IASF/CNR, Via Piero Gobetti 101, 40129 Bologna, Italy
      \and Service d'Astrophysique, DAPNIA/DSM/CEA, 91191 Gif-sur-Yvette, France}

\date{Received / Accepted}

\abstract{The discovery of the X-ray source \IGR\ by \INTEGRAL\ was reported on
          9 February 2004. Regular
          monitoring by \INTEGRAL\ shows that \IGR\ is a persistent hard X-ray source with
          an average count rate of 0.96 \cps\ ($\sim$6.4 mCrab)
          in the 20--60 keV energy band. A follow-up observation with \XMM\, which  was
          performed on 21 March 21 2004, showed that the source is located at
          R.A. (2000.0) $=17^{h}25^{m}11.4^{s}$ and Dec. $= -36\degr 16\arcmin 58.6\arcsec$
          with an uncertainty of $4\arcsec$. The only infra-red counterpart to
          be found within the \XMM\ error circle was \object{2MASS~J17251139$-$3616575},
          which has a Ks-band magnitude of 10.7 and is located 1$\arcsec$ away
          from the \XMM\ position.

          The analysis of the combined \INTEGRAL\ and \XMM\
          observations shows that the source is a binary X-ray pulsar with a spin period
          of 413.7~s and an orbital period of 9.72~days. The spectrum can be
          fitted with a flat power law plus an
          energy cut off ($\Gamma \sim 0.02,E_{\mathrm{c}} \sim 8.2\,\mathrm{keV}$)
          or a Comptonized model ($kT_{\tty{e}}\sim 5.5\,\mathrm{keV}, \tau\sim 7.8$).
          The spectrum also indicates a large hydrogen column density of
          $N_{\mathrm{H}}\sim 15\,10^{22}\,\mathrm{atoms}\,\unit{cm}{-2}$
          suggesting an intrinsic absorption. The Fe K$\alpha$ line at 6.4 keV
          is clearly detected. Phase-resolved spectroscopy does not show any
          variation in the continuum except the total emitted flux. The
          absorption is constant along the pulse phase.
          This source can be associated with \object{EXO~1722$-$363} as
          both systems show common timing and spectral features.
	  The observations suggest that
          the source is a wind-fed accreting pulsar accompanied by a supergiant star.
          
}

\maketitle

%

\section{Introduction}

X-ray binaries consist of a compact object, either a black hole or
a neutron star, accreting matter from a companion star; they are usually
classified according to the mass of the companion as a high mass (HMXB,
$M_{\mathrm{C}}\ga 10\,$\Msol), intermediate-mass (IMXB, $M_{\mathrm{C}}=
1$--$10\,$\Msol) or low mass X-ray binary (LMXB, $M_{\mathrm{C}}\la 1\,$\Msol).
The LMXB and IMXB accrete matter through Roche-Lobe overflow from the companion
star and through an accretion disk around the compact object
\citep{TaurisHeuvel05}.

The HMXB can be divided into two categories as the companion star can be
either an OB supergiant or a Be star.  OB supergiant stars feed the compact
object through strong stellar winds and/or in some cases Roche-Lobe overflow. Be
stars expell matter around their equator that fuels the compact object. The
orbits of Be binary systems are eccentric and generally have a longer period
than OB supergiant systems ($\ga 15$ days). The X-ray emission of HMXB presents
a wide variety of patterns: from transient to persistent, outbursts on different
times scales (seconds, days and/or years), periodic modulations, eclipses and
others. Strongly magnetized neutron stars accompanied by massive stars show
periodic pulsations; the accreted matter is funneled towards the poles
by the magnetic field leading to an increase of the observed X-ray emission when
these regions cross the line of sight. These HMXB usually show a hard X-ray
spectrum between 2--10 keV with an energy cutoff around 10 keV.

Several new hard X-ray sources have been discovered by \INTEGRAL\
\citep{Winkleral03} in surveys of the galactic plane. Most of them have been
detected by IBIS/ISGRI \citep{Ubertinial03,Lebrunal03}, the most sensitive
instrument on board \INTEGRAL\ between 20 and 300 keV.
The brightest sources detected during the first year of the \INTEGRAL\ mission
are listed in \citet{Birdal04}. A few tens of them have never been detected
before \INTEGRAL's observation.
Most of these new objects show common features in their spectra, such as a high
intrinsic low-energy absorption, and they are believed to be HMXB
\citep{Walteral03,Rodriguezal03,Patelal04}.

\IGR's discovery was reported on February 9, 2004, with 13 other 
hard X-ray sources \citep{Walteral04} detected with ISGRI in all-sky mosaic
images built from core programme data.
However, there is evidence that \IGR\ has already been observed with previous
missions. A galactic plane scan performed by \emph{EXOSAT}\ in June 1984 revealed a point
like X-ray source emission, GPS 1722-363 \citep{Warwickal88}, at a position not
compatible with \IGR\ but with a low accuracy. \emph{Ginga} observations in 1987 and
1988 confirmed the presence of a powerful X-ray accretor neutron star, X1722-36,
with a pulsation of 413.9 s and important variations of the intensity in X-rays
\citep{Tawaraal89}. The source also showed a hard spectrum with important
low-energy absorption and an emission line at 6.2~keV. Further investigations
with Ginga in 1988 confirmed the spectral analysis, and lower limits for the
orbital period of 9 days and the mass of the primary star of 15~\Msol were
deduced from pulse timing analysis \citep{Takeuchial90}. Both investigations
conclude that the system is a HMXB.
Recently, those results were confirmed by \citet{Corbetal05} using 
\emph{RXTE} data. They also found an orbital period of 9.741$\pm$0.004 d.

\IGR\ is regularly monitored by \INTEGRAL. A follow-up observation lasting three
hours was performed with the X-ray Multi-Mirror Mission (\XMM) on 21 March 
2004. The new data available on this source is presented in this paper.
\INTEGRAL\ and \XMM\ observations and data analysis techniques are described in
Sects.~\ref{SecObs} and \ref{SecAnalysis}, respectively. The results are
presented in Sect.~\ref{SecResults} and discussed in Sect.~\ref{SecDisc}. 
The conclusions are presented in Sect.~\ref{SecConc}.


\section{Observations}\label{SecObs}

\subsection{\INTEGRAL}\label{SubINTEGRAL}

\INTEGRAL\ is a hard X-ray and $\gamma$-ray observatory of the European
Space Agency (ESA) launched on 17 October 2002. The payload consists of four
instruments: the imager IBIS with two detector layers, ISGRI (20 keV--1 MeV)
and PICsIT (200 keV--10 MeV, \citet{Labantial2003}); the spectrometer SPI 
(20 keV--8 MeV, \citet{Vedrenneal03}); the X-ray monitor JEM-X (3--30 keV,
\citet{Lundal03}) and the optical camera OMC (V filter, \citet{Mas-Hesseal03}).
Most of the \INTEGRAL\ observing time is spent in the Galactic Plane.

\IGR\ is located close to the Galactic Centre, which has been observed regularly.
The observing strategy consists of pointings each lasting $\sim$~30~minutes
distributed in various grids around the Galactic Plane. The focus of the effort
for this paper is on IBIS/ISGRI, since IBIS/PICsIT and SPI are less sensitive at
energies lower than 300 keV where the source is detected.
We did not use data from the JEM-X instrument because its smaller field of view and the
dithering observation approach of \INTEGRAL\ mean that the effective exposure on \IGR\ is ten
times smaller than obtained with ISGRI. 

The data set consists of core programme data obtained until MJD~53341.1 and of
public data obtained until MJD~52928.3, giving a total exposure of 6.5~Ms.
Table~\ref{tabDataSet} lists the source visibility periods, and for
each of them the fraction of the time when the source was effectively
observed in the partially-coded field of view (PCFOV).
\begin{table*}
\caption{The \INTEGRAL\ Data set. We selected all public and core programme revolutions
         when \IGR\ was within the ISGRI PCFOV of $29\degr \times 29 \degr$.
         The observing time fraction was calculated as the ratio between the
         net exposure time on the source and the elapsed time during the visibility
         periods.
	}
\label{tabDataSet}
\centering
\begin{tabular}{c c c c c}
\hline\hline
Visibility Periods & Start & End   & Revolutions & Observing time fraction \\ 
     & [MJD] & [MJD] & & \\ 
\hline			    
   1 & 52671 & 52752 &  37- 63 & 0.23\\
   2 & 52859 & 52925 & 100-121 & 0.51\\
   3 & 53050 & 53116 & 164-185 & 0.20\\
   4 & 53236 & 53294 & 226-244 & 0.16\\
\hline
\end{tabular}
\end{table*}

\subsection{\XMM}\label{SubXMM}

\IGR\ was observed by \XMM\ \citep{Jansenal01} on 21~March~2004, from 13:02:45
to 16:04:45 UTC (MJD 53085.544--53085.671) for a total exposure of 11 ks. \XMM\
operates with three instruments that cover the optical/UV and X-ray spectral
bands. The main instrument for the purpose of this work is the European Photon Imaging Camera (EPIC)
instrument, which consists of two MOS \citep{Turneral01} and one pn
\citep{Struderal01} CCD cameras.  EPIC has a 30$\arcmin$ field of
view, and coverage of the 0.15--12~keV energy range with imaging, timing and
spectral capabilities. The EPIC/MOS[12] and pn were all operating in imaging
science mode with a large window and a medium filter.

The \XMM\ observation was simultaneous with three \INTEGRAL\ pointings in
revolution 175 between MJD~53085.542 and 53085.667. However, the signal to noise
(S/N) of ISGRI data during those simultaneous observations was too low to
perform cross analysis.


\section{Data Analysis}\label{SecAnalysis}

\subsection{\INTEGRAL}\label{SubINTEGRAL}

The data were reduced with the Offline Scientific Analysis version 4.2 software
(OSA 4.2) that is publicly released by the \emph{INTEGRAL} Science Data Centre
(ISDC) \citep{Courvoisieral03}.  All the pointings for which the
source was either within the ISGRI fully coded (FCFOV, $9\degr \times 9 \degr$)
or partially coded (PCFOV, $29\degr \times 29 \degr$) fields of view were analysed. Around
3000 pointings, distributed between revolutions 37 and 244, were selected.
Sky images were extracted for each pointing and combined into mosaic images with longer
exposures. As the source is weak, light curves were extracted from the imaging
results (pointings and mosaics). During flares, light curves were extracted on
shorter time scales using {\tt ii\_light} v7.3. Extracting high-energy spectra
with OSA 4.2 remains a difficult task for faint sources.  Spectra were extracted 
using {\tt ii\_spectra\_extract} v2.3.1 and from the mosaic images, and compared.
In the energy range in which the source is detected, both methods give consistent results. 
The redistribution matrix and ancillary response files (RMF and ARF) used were
{\tt isgr\_rmf\_grp\_0012.fits} and {\tt isgr\_arf\_rsp\_0006.fits},
respectively. The RMF was rebinned in to 17 channels.
Average fluxes and the source position were obtained from the mosaic
images with {\tt mosaic\_spec} v1.0\footnote{{\tt ii\_light} and 
{\tt ii\_spectra\_extract} are OSA 4.2 standard tools.
{\tt mosaic\_spec} is released with OSA 5.0.}.

\subsection{\XMM}\label{SubXMM}

The Science Analysis System (SAS) version~6.1.0 was used to produce new event
lists for the EPIC instrument running {\tt epchain} for pn and {\tt emchain}
for MOS[12]. The event lists were corrected for enhanced background features
at energies higher than 10~keV, disregarding time lapses when count rates above
10 keV exceeded 4 \cps\ for pn, 1 count$\,\unit{s}{-1}$ for MOS1 and 
1 count$\,\unit{s}{-1}$ for MOS2. Finally, only 5.9 ks of 9.2 ks, 6.2 ks of 10.7 ks
and 7.0 ks of 10.7 ks for each instrument, respectively, were kept as good time
intervals.

Images were built from the cleaned event lists for the MOS[12] and pn cameras
with 2$\arcsec$ and 4$\arcsec$ resolution, respectively. Bad pixels were
disregarded and good events were selected until the quadruple level. The lower
threshold was fixed at 0.8~keV as recommended in the calibration status
documentation (see calibration document XMM-SOC-CAL-TN-0018\footnote{
{\tt http://xmm.vilspa.esa.es/external/xmm\_sw\_cal/\newline
calib/documentation/index.shtml\#EPIC}}
, p25).

To find the source location, the SAS task {\tt edetect\_chain} was used on each
individual EPIC camera.  Four images, with energy ranges of 0.5--2,
2--4.5, 4.5--7.5 and 7.5--12~keV, were created. The source position was
calculated as the mean of the best position from each MOS and pn camera.

The source was clearly visible in the pn images in CCD~1, near the read-out node.
An event list of source$+$background counts was selected from a circle of 50$\arcsec$
around the bright object. The background was estimated from a region adjacent to the source region,
with a similar size and at the same distance from the read-out node in the same CCD.
To build light curves, only single and double events with an energy between 0.4 and 10~keV were collected,
as advocated in the user guide. The background light curve was subtracted using
the {\tt FTOOLS lcmath}.

Source$+$background and background spectra were extracted from all EPIC cameras,
MOS[12] and pn, disregarding bad pixels and selecting single and double events.
Specific redistribution matrix and ancillary response files were generated
for each EPIC instrument with the standard tasks {\tt rmfgen} and {\tt arfgen},
respectively. An average spectrum of the source from the complete observation
was first obtained and the {\tt Xspec} version 11.3.1 package was used to fit
and plot the resulting spectra corrected for the background. 

For the phase-resolved spectra, the {\tt Xselect} version 2.2 software was used to
select events corresponding to a specific phase bin given the start epoch and
the periodicity of the object. Source and background spectra for each
selected phase bin were calculated from the same regions as defined previously.


\section{Results}\label{SecResults}

\subsection{Source position}\label{SubIma}

\begin{figure}
  \centering
  \resizebox{\hsize}{!}{\includegraphics[width=0.3\textwidth]{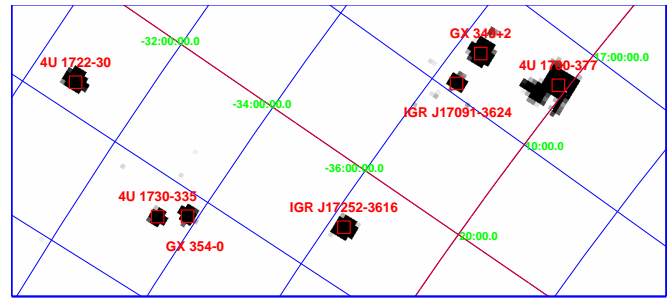}}
  \caption{Mosaic image of \IGR\ with ISGRI during revolution 106 (MJD 52877.4--52880.4).
          }
  \label{FigIsgriMosImgRevol0106}
\end{figure}

The source's brightest epoch occurred during revolution 106 between MJD 52877.4
and 52880.4 with an average flux of 4.08$\pm$0.09 \cps\ and a significance of
47$\sigma$.  The ISGRI 20--60~keV mosaic of that epoch was used to extract the hard X-ray source
position at R.A. (2000.0) $= 17^{h}25^{m}10^{s}$ and Dec. $=-36\degr 17\arcmin 18\arcsec$,
with an uncertainty of $23\arcsec$ that includes a $10\arcsec$ systematic error
due to the instrument misalignment (see Fig.~\ref{FigIsgriMosImgRevol0106}).
This statistical error is of the same order as the systematic error that
comes from the image reconstruction, which is proportional to the source
significance: $\sim 36 \arcsec$ for a S/N of 47 \citep{Grosal03}.
This position is compatible with the position obtained by \citet{Walteral04}.

\begin{figure}
  \centering
  \resizebox{\hsize}{!}{\includegraphics[width=0.2\textwidth]{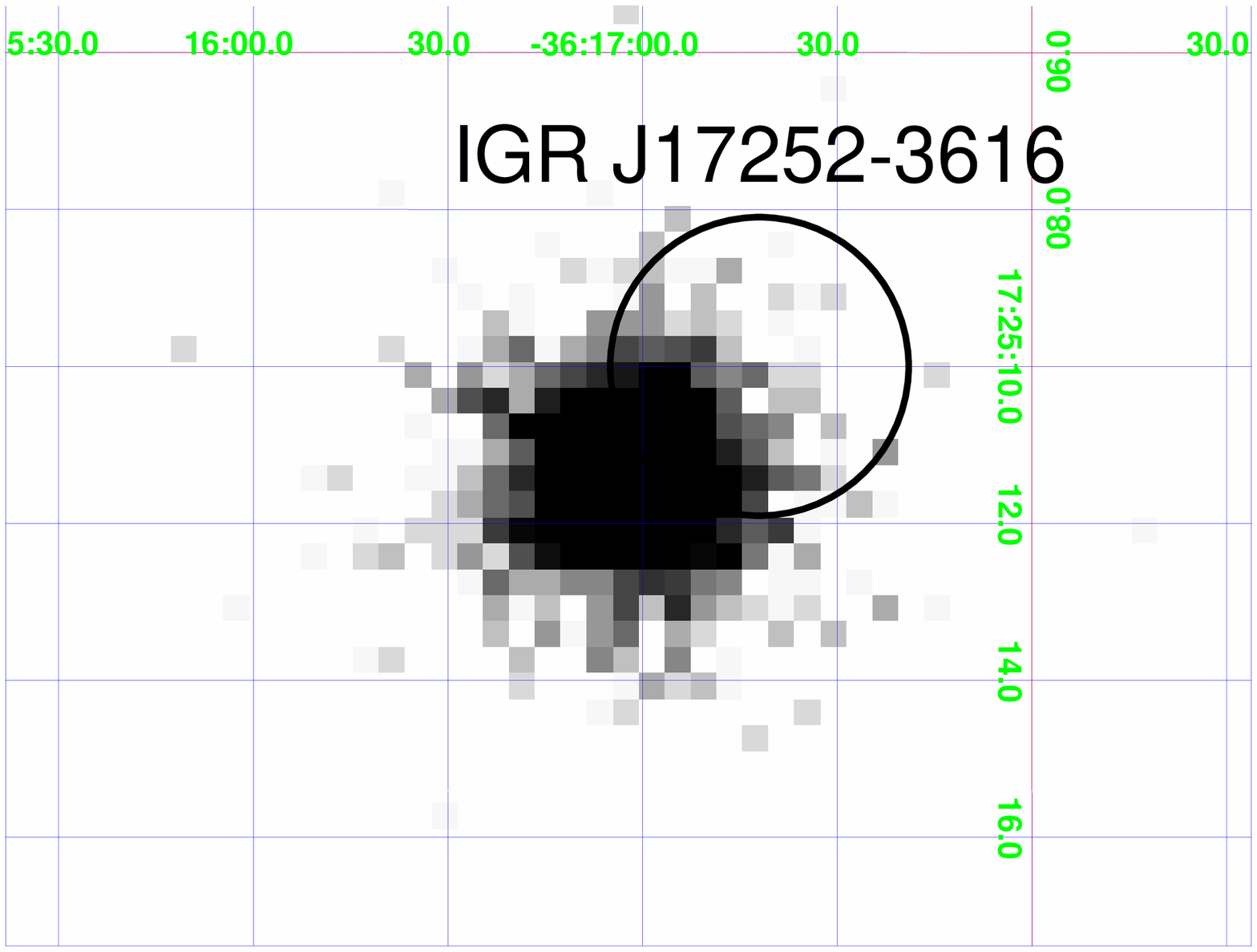}}
  \caption{\IGR\ EPIC/MOS1 image with the $23\arcsec$ \INTEGRAL/ISGRI error circle
          }
  \label{FigMOS1Img}
\end{figure}

Fig.~\ref{FigMOS1Img} shows the \XMM\ EPIC/MOS1 image with the ISGRI error circle.
A single X-ray source corresponds to the \INTEGRAL\ position. Its position is
R.A. (2000.0) $= 17^{h}25^{m}11.4^{s}$ and Dec. $= -36\degr 16\arcmin 58.6\arcsec$
with an uncertainty of $4\arcsec$.

\begin{figure}
  \centering
  \resizebox{\hsize}{!}{\includegraphics[width=0.2\textwidth]{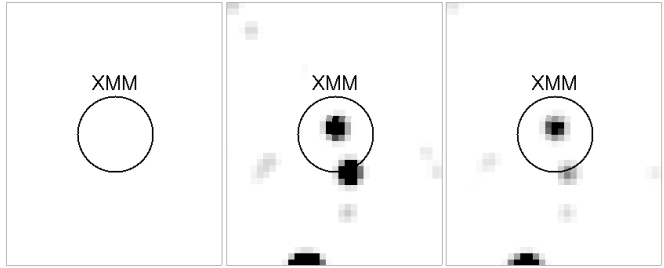}}
  \caption{Infrared counterpart of \IGR. Images taken from the 2MASS survey:
           J-band (left), H-band (middle) and K-band (right). The EPIC error
           circle of 4$\arcsec$ is displayed. \object{2MASS J17251139-3616575} is only
           present in the H and K images.
          }
  \label{Fig2MASSImg}
\end{figure}

With the accurate source position obtained with EPIC, the Two Microns All-Sky
Survey (2MASS) catalog \citep{Cutrial03} was searched for an infra-red counterpart.
Only one 2MASS source appears fully within the 4$\arcsec$ EPIC error circle,
\object{2MASS J17251139-3616575}, located 1$\arcsec$ away from the EPIC position.
The source is not detected in the J-band with an upper limit of 14.2~magn at
95\% confidence level, but appears in the H-band with 11.8~magn and the
K$_{\mathrm{s}}$-band with 10.7~magn (see Fig.~\ref{Fig2MASSImg}).
The colours obtained when dereddening those infra-red magnitudes can either
be interpreted as a close cool star or a distant hot star \citep{Walteral05}.

\subsection{Timing analysis}\label{SubTime}

\subsubsection{Long-term variability}

\begin{figure}
  \centering
  \resizebox{\hsize}{!}{\includegraphics[]{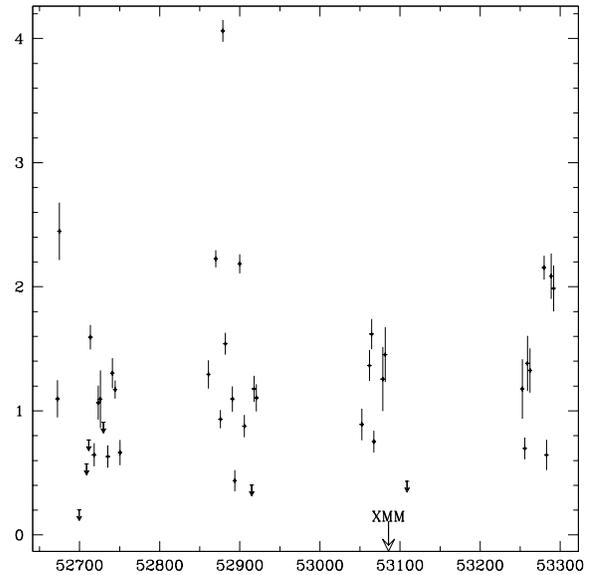}}
  \caption{\IGR\ 20--60 keV light curve with average count rate per revolution
           ($\sim$3 days).
	   3$\sigma$ upper limits are indicated when the source was not
	   detected. The \XMM\ observation time is shown.
          }
  \label{FigIsgrilcRevol}
\end{figure}

The long term variability of \IGR\ was studied using ISGRI data in two
energy ranges: 20--60 keV and 60--150 keV. Mosaic images were built for each
three days revolution and light curves extracted for both energy bands. When
the source was not detected, 3$\sigma$ flux upper limits were calculated. Net
exposures higher than 5~ks at the source location were considered. Typical net
exposures for one revolution vary from a few~ks to several tens of ks. The
source was never detected above 60 keV. 
The 3$\sigma$ 60--150 keV average count rate upper limit is $\sim$0.3 \cps\
($\sim$7 mCrab) with the highest and lowest values being $\sim$0.6 \cps\
($\sim$14 mCrab) and $\sim$0.6 \cps\ ($\sim$3.5 mCrab).
The 20--60 keV light curve on the time
scale of a spacecraft revolution is shown in Fig.~\ref{FigIsgrilcRevol}.

The source is detected in almost every single revolution.
Its average 20--60~keV flux is 0.96$\pm$0.01 \cps\ ($\sim$6.4 mCrab),
derived from a 20--60~keV mosaic image generated with data of the first three
visibility periods for a total net exposure of 1.7 Ms. The average flux of the
object selecting only revolutions when the source is detected is 1.39$\pm$0.02 \cps\
($\sim$9.3 mCrab) as calculated from Fig.~\ref{FigIsgrilcRevol}. When selecting
revolutions for which only upper limits are available (rev. 46, 49, 50, 56, 118
and 183) and building a mosaic image with a net exposure of 165 ks, the source
is detected with a significance of 5$\sigma$, and the mean 20--60 keV flux reaches 
0.19$\pm$0.04 \cps\ ($\sim$1.3 mCrab); this is five times less than the average
mean flux over all the revolutions and seven times less than the average flux
over revolutions when the source was detected.
Nevertheless, the source is persistent as, once cumulating enough exposure
time, it remains detectable.

On revolution time scales, the source flared up to 4 \cps\ ($\sim$27 mCrab),
or a factor of 10 stronger than the lowest detection of 0.4 \cps\ ($\sim$2.7 mCrab),
on one occasion (see Fig.~\ref{FigIsgrilcRevol}). 
\IGR\ varies by a factor $\sim$4 outside of this flare.

\begin{figure}
  \centering
  \resizebox{\hsize}{!}{\includegraphics{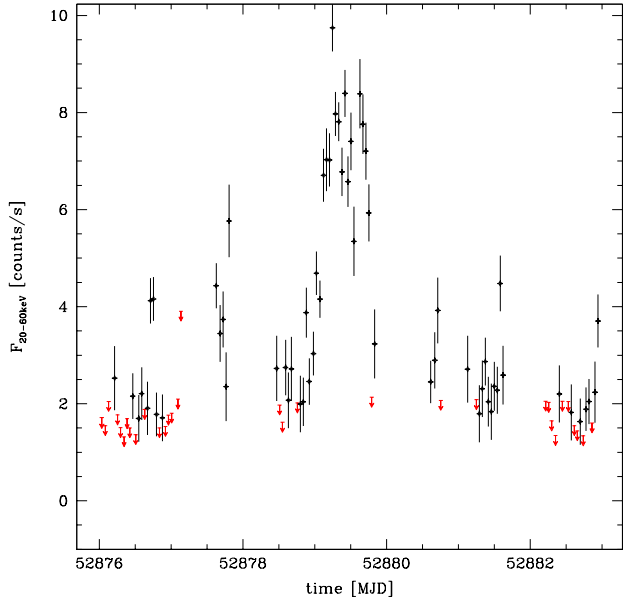}}
  \caption{\IGR\ 20--60 keV light curve with average count rates per pointing ($\sim$30 min).
           Around MJD 52879, we observe the brightest flare during the four
           visibility periods. 3$\sigma$ upper limits are indicated when the
           source was not detected.
          }
  \label{FiglcZoomIsgri}
\end{figure}

Variability on time scales shorter than a single revolution was also investigated.
Light curves based on flux per pointing were built considering the full
\INTEGRAL\ data set in the 20--60 keV energy range. The source is often not
detected in single pointings. The average count rate per pointing when the
source is detected with a significance higher than 4$\sigma$ varies between 2 and 4
\cps\ ($\sim$13--27 mCrab) outside the flares. Four flares are detected in
total, three in the 2nd visibility period and one in the 4th visibility period.
The brightest flare occured at MJD 52879 when the flux increased from $\sim$2
\cps\ to a peak of 10 \cps\ ($\sim$67 mCrab) and decreased again in $\sim$1 day
(see Fig.~\ref{FiglcZoomIsgri}). The other flares reached a flux level of
$\sim$6--8 \cps\ ($\sim$40--53 mCrab) and lasted less than one day.
Considering the observed count rates limit values of 0.2 and 10 \cps, the
source varies by a factor of more than 50 between different epochs.

\subsubsection{Pulse period}

\begin{figure}
  \centering
  \resizebox{\hsize}{!}{\includegraphics{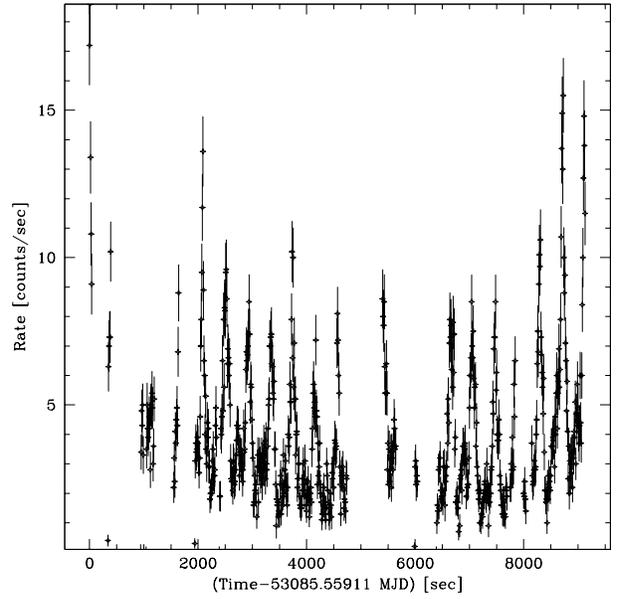}}
  \caption{\IGR\ EPIC/pn 10 s binned light curve. The gaps correspond to the
           periods with enhanced background activity that were discarded before
           starting the analysis.
          }
  \label{Figpnlc}
\end{figure}

Fig.~\ref{Figpnlc} shows the \XMM\ light curve. 
A periodic oscillation of $\sim$~400~s is clearly visible.
A power spectral density distribution was created from the entire light curve
in order to search for periodicity. A main peak was found around $\sim$400~s with a
harmonic at 200~s.
 
\begin{figure}
  \centering
  \resizebox{\hsize}{!}{\includegraphics{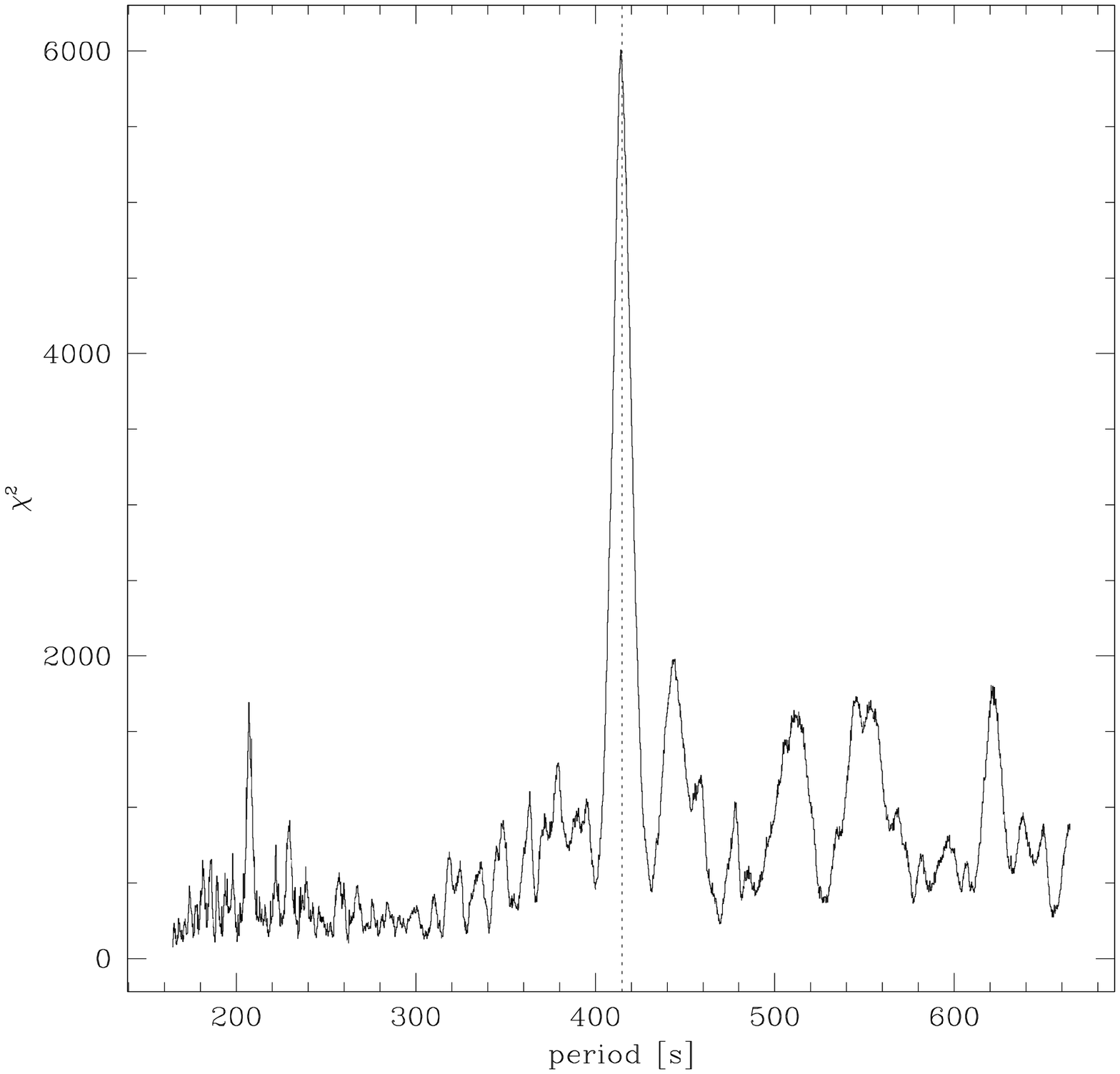}} 
  \resizebox{\hsize}{!}{\includegraphics{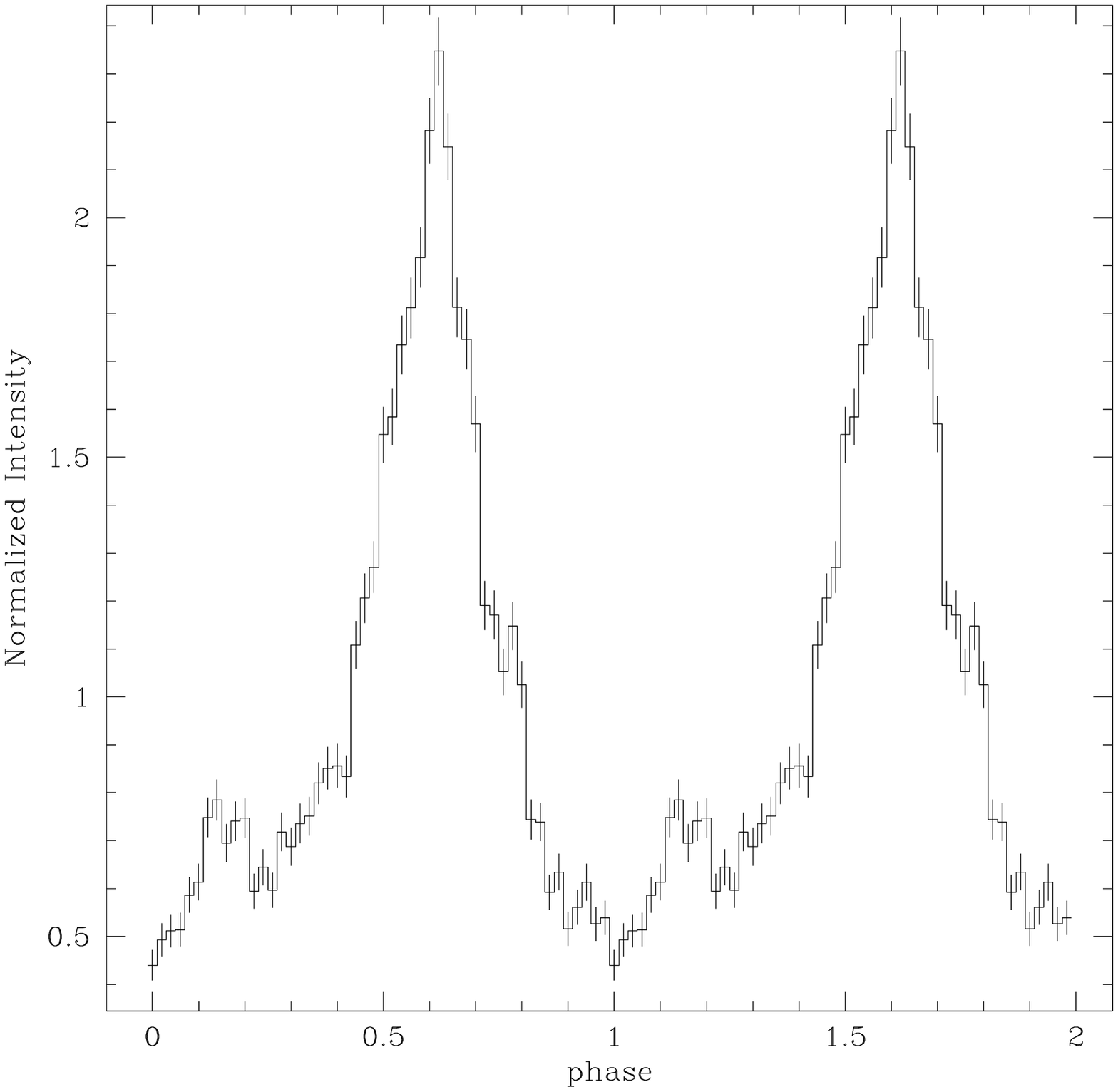}}
  \caption{\textit{top:} $\chi^{2}$ distribution of trials periods for epoch folding search on
           EPIC/pn data. The best value was fitted with a Gaussian. The best
           period of 414.8$\pm$0.5 s is indicated with a dashed-line.
           \textit{bottom:} \IGR\ pulse phase folded light curve in the 0.4--10 keV energy band
           obtained with the best spin period. The zero epoch is MJD 53085.55144.
          }
  \label{FigpnFes}
\end{figure}

Starting from the estimated period, the best period was searched for in the
$\chi^{2}$ distribution when folding the light curve over a range of different
periods. A significant peak was obtained, giving the best period of
414.8$\pm$0.5~s when fitted with a Gaussian (see Fig.~\ref{FigpnFes} \textit{top}). 
The uncertainty on the period was estimated using Eq. 14 of
\citet{HorneBaliunas86} and Eq. 2 of \citet{Hillal05}. Lomb-Scargle
periodograms were generated using \citet{PressRybicki89} fast method.
The light curve was folded with the best period to
obtain the pulse profile between 0.4--10 keV (see Fig.\ref{FigpnFes} \textit{bottom}).
The \XMM\ folded light curve is defined so that the minimum flux is set at phase
0 corresponding to MJD 53085.55144. A first local maximum is observed at phase $\sim
0.15$ followed by a broad maximum at phase $\sim$0.6.
The pulse fraction, defined as  
$P_{\mathrm{f}}=(I_{\mathrm{max}}-I_{\mathrm{min}})/(I_{\mathrm{max}}+I_{\mathrm{min}})$
with $I_{\mathrm{max}}$ and $I_{\mathrm{min}}$ being the maximum and minimum
intensities of the folded light curve, respectively, reaches $68\pm 3$\%.

\begin{figure}
  \centering
  \resizebox{\hsize}{!}{\includegraphics{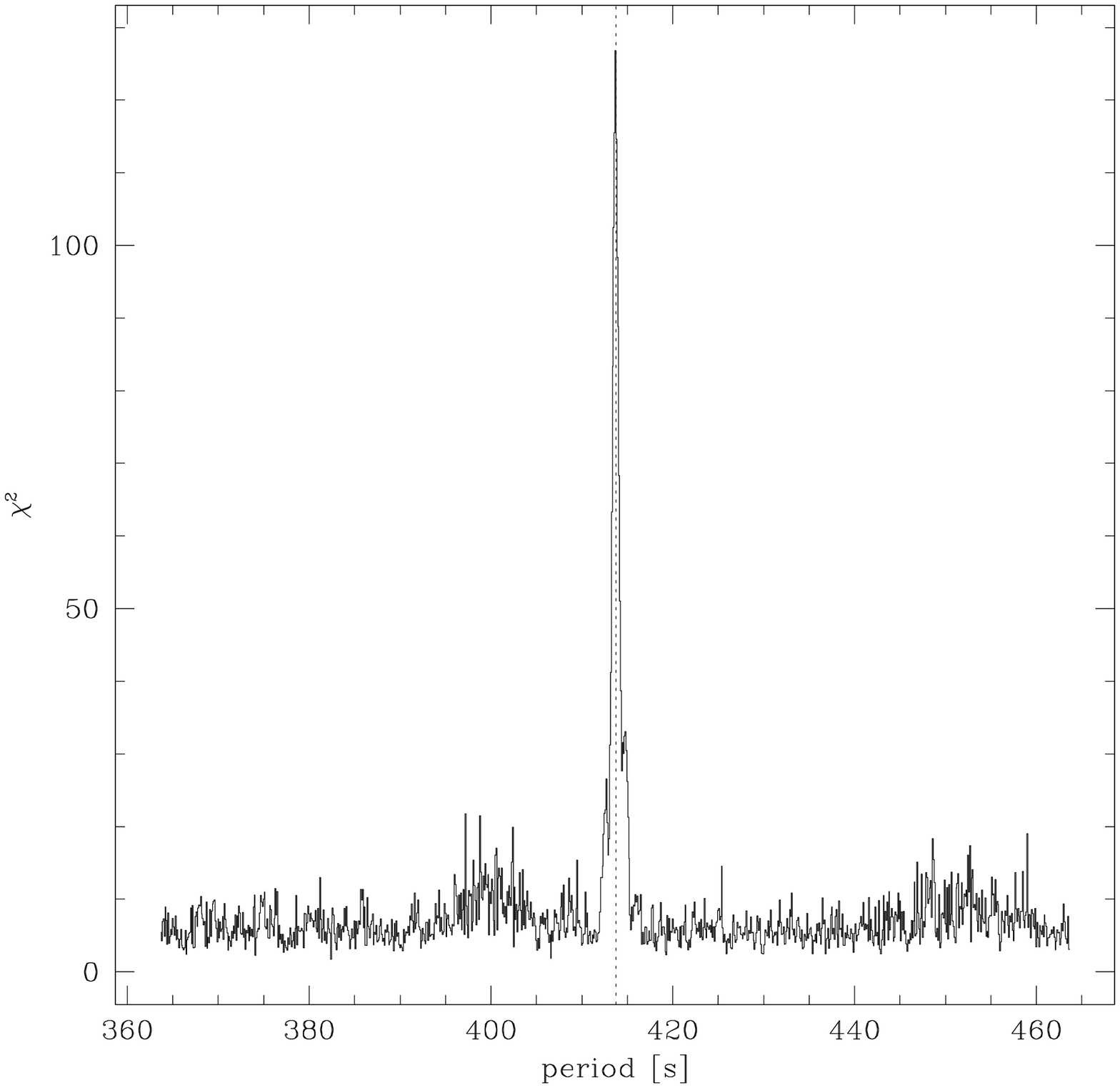}}
  \resizebox{\hsize}{!}{\includegraphics{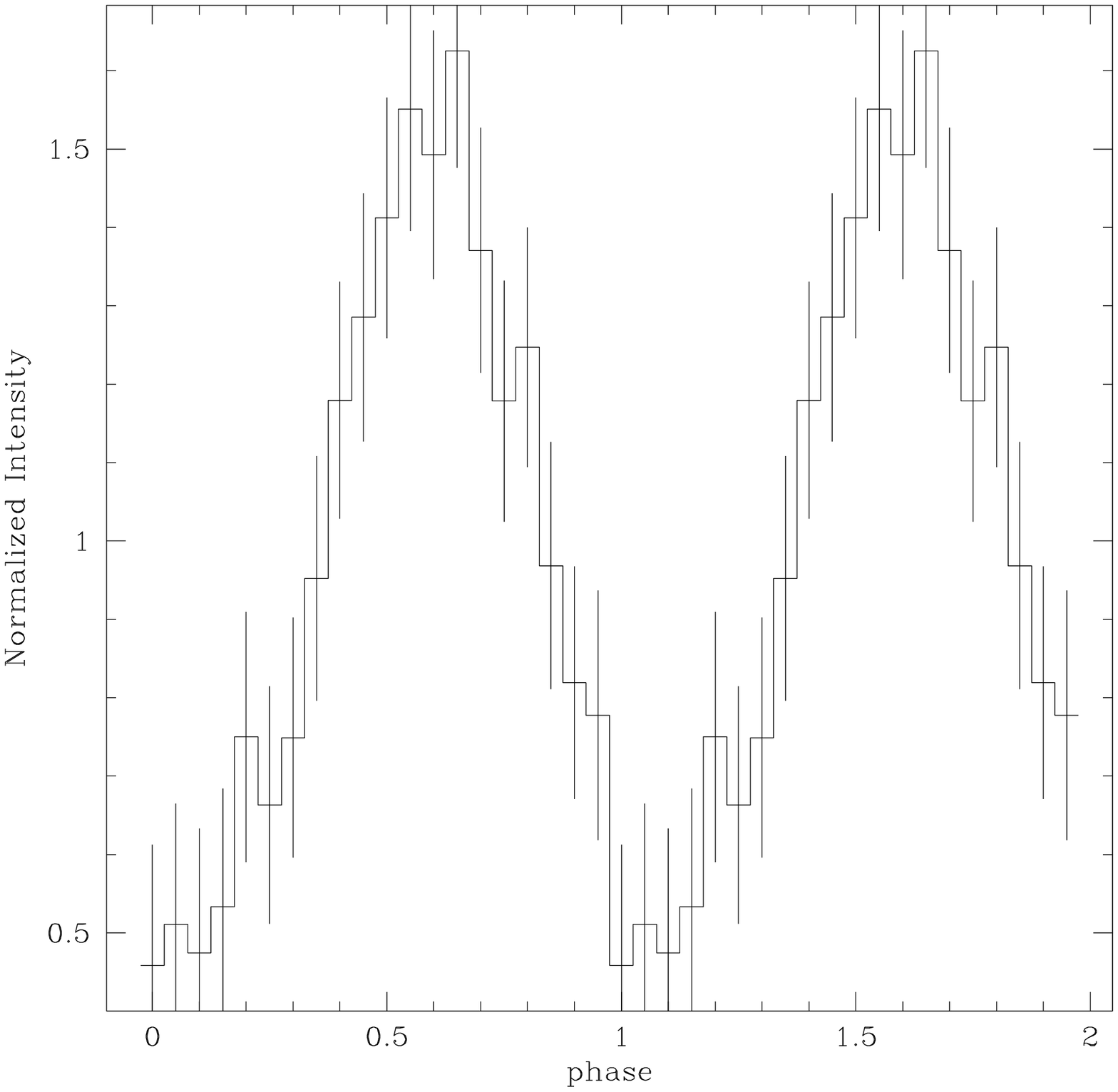}}
  \caption{\textit{top:} \IGR\ IBIS/ISGRI epoch folding search $\chi^{2}$ distribution. The
           best period was fitted with a Gaussian and resulted in
           413.7$\pm$0.3~s (see dashed-line).
           \textit{bottom:} ISGRI folded light curve built with data from revolution 106
           (MJD 52877.4--52880.4). \IGR\ pulse phase folded light curve in the
           20--40 keV energy band was obtained with the best spin period.
           The zero epoch is MJD 52877.50072.
          }
  \label{FigIsgriFes}
\end{figure}

An ISGRI 50~s binned light curve was made during the bright flare observed
at MJD 52879 in two energy bands 20--40 keV and 40--60 keV. Epoch folding the
\INTEGRAL\ data, starting with the period found in the EPIC/pn light curve,
returned a period of 413.7$\pm$0.3~s in the 20--40 keV band
(see~Fig.\ref{FigIsgriFes} \textit{top}). The signal to noise of the 40--60 keV
light curve is not significant enough to detect the modulation. The pulse
profile shape shows a broad peak without complex
structures (see Fig.~\ref{FigIsgriFes} \textit{bottom}).
The pulse fraction reaches $56\pm$12\% and is consistent with the value observed
between 0.4--10 keV.

The shape of the X- and $\gamma$-ray pulse profiles show some differences. Due to
the less significant detection in the ISGRI data, the number of phase bins is
different in the pn and ISGRI folded light curves. The main peak in ISGRI seems
broader than in pn. There is also a hint of a secondary peak in the pn data that
is absent in the ISGRI data. The folded light curves cannot be compared in
phase, because the phase 0 of both pulse profiles do not correspond in time as
the observations are separated by 7 months. Notice that the 20--60 keV source
intensity was three times brighter during the flare (MJD 52879) than during
the \XMM\ observation ($\sim$MJD 53085.6) (see Fig.~\ref{FigIsgrilcRevol}).

\subsubsection{Orbital period}

\begin{figure}
  \centering
  \resizebox{\hsize}{!}{\includegraphics{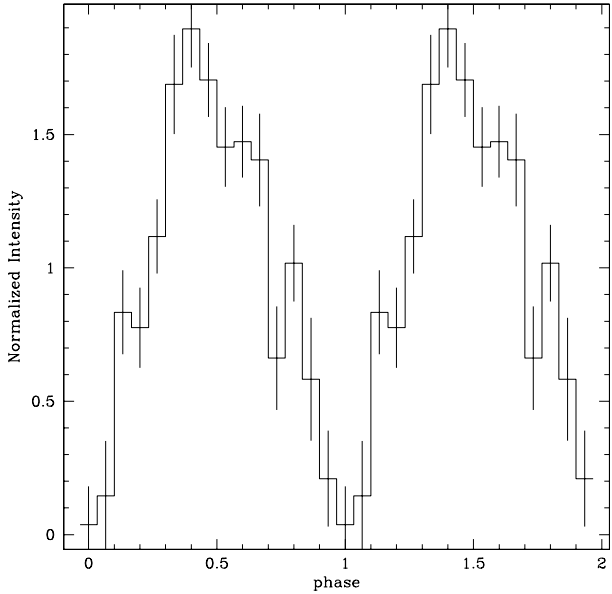}}
  \caption{\IGR\ orbital folded light curve in the 20--60 keV energy band was
           obtained with the best period of 9.72~d. The zero epoch is MJD 52671.
           The \XMM\ observation correspond to phase $\sim$0.7.
          }
  \label{FigIsgriOrbPhase}
\end{figure}

Indications of an orbital period in the 20--60~keV light curve were searched for, 
based on flux per pointing. A Lomb-Scargle periodogram was generated between 4
and 28~days. The periods with flaring activity were removed in the light curve
before searching for any coherent modulation.
A significant peak at 9.72$\pm$0.09~days is visible and the folded
light curve is shown in Fig.~\ref{FigIsgriOrbPhase}. A minimum flux consistent
with an eclipse is visible at phase 0. It lasts between phases 0.97--1.1 that
corresponds to a lapse of time of $\sim$1.26 days. The normalized intensity
smoothly increases from phase 0.1 to 0.3 to reach a plateau that lasts
$\sim$3.9~days. Afterwards, the intensity gradually decreases from phase 0.7 to the
start of the eclipse at phase 0.97. The \XMM\ observation corresponds to phase
$\sim$0.7.

\subsection{Spectral analysis}\label{SecSpectra}

\begin{figure}
  \centering
  \resizebox{\hsize}{!}{\includegraphics[angle=-90]{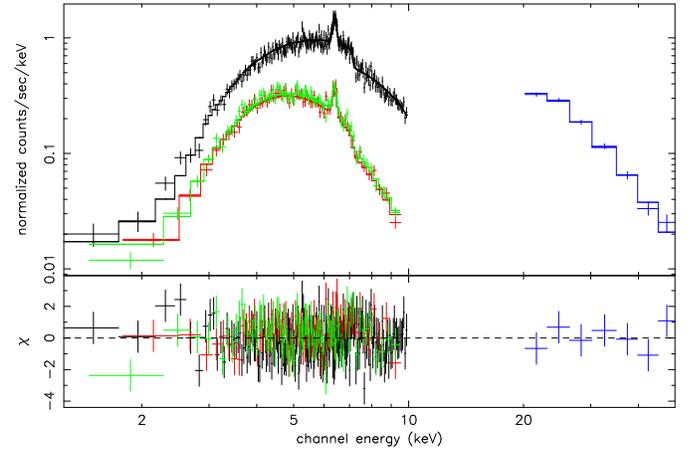}}
  \caption{Combined \IGR\ EPIC+ISGRI averaged spectrum. All three EPIC spectra
           are displayed in the X-ray band: pn (top) and MOS[12] (bottom). The
           ISGRI spectra is in the hard X-ray band. The source is detected up to
           50 keV. CTT model is used.
          }
  \label{FigXMMSpec}
\end{figure}

The X-ray spectral bins were grouped to have at least 100 counts per channel
(see Fig.~\ref{FigXMMSpec}). This allows the use of the $\chi^{2}$ statistic.
The spectrum is strongly absorbed at energies lower than 4~keV.
An emission line at 6.4~keV and an edge at 7~keV are clearly detected. The
spectra were first fitted with a simple absorbed power law and a blackbody
model. In terms of reduced $\chi^{2}$, these simple models fitted the \XMM\ data
well. In order to distinguish between the two models, the EPIC and ISGRI spectra
are used together.

The ISGRI spectrum was extracted from revolution 106 data when the source intensity
was the strongest (see Fig.~\ref{FiglcZoomIsgri}) and also from an average mosaic
image of all visibility periods. The two spectra do not show any significant difference
beside their normalisations. In the following spectral analysis, the
revolution 106 source spectrum was used since it has the highest S/N.

When fitting EPIC and ISGRI spectra together, an inter calibration
factor, $C_{\tty{ISGRI}}$, was applied to take into account the different source flux
levels during the non simultaneous observations. The simple absorbed power law and
blackbody models failed to fit the data between 1--50~keV. The blackbody is too
narrow and the power law requires a high-energy cutoff.

A cutoff power law (CPL) was therefore chosen as a phenomenological model to
fit the data. Moreover, a more physical model based on Comptonization (CTT)
\citep{Titarchuk94} was also adopted as suggested by other observations
\citep{Torrejonal04,Bodagheeal05,Masettial05,Walteral05}. In {\tt Xspec}, the
CPL model is defined as {\tt cst*wabs(bbody+vphabs(cutoffpl+ga))} and
CTT as {\tt cst*wabs(bbody+vphabs(compTT+ga))}.
The 6.4~keV line was fitted with a Gaussian. An absorption model was used, where
the iron abundance was left as a free parameter. The soft excess at low energies
was modelled with a black body. The galactic absorption is also taken into
account.
All the spectral parameters errors were calculated at the 
90\% confidence level. Parameters resulting from the spectral fitting are listed
in Table~\ref{TabSpec}.

The CPL model gives typical values of an accreting pulsar with a flat powerlaw,
$\Gamma=0.0\pm 0.1$, and an energy cutoff, $E_{\mathrm{c}}=8.2\pm 0.4$ keV. The
CTT model also fits the data well with characteristic values for the electron
temperature of $kT_{\mathrm{e}}=5.5\pm 0.2$ keV and an optical depth of 
$\tau=7.8\pm 0.6$. In both cases, the unabsorbed flux between 2--12 keV and
13--100 keV is 2 and 5 $10^{-10}\,\mathrm{ergs}\,\unit{cm}{-2}\,\unit{s}{-1}$,
respectively. 

The line centroid at $6.40\pm 0.01$ keV is compatible with what is expected from
cold iron. The line width is consistent with zero and was fixed for the spectral
fits. The equivalent width of that line is 84~eV and was calculated considering
the unabsorbed continuum.

A soft excess is required below 3~keV and was represented by a blackbody
absorbed by the galactic column density (without soft excess, 
$\chi^{2}/\mathrm{d.o.f.}$=415/376 and 420/376 for CPL and CTT models,
respectively). In this case, the 1-2~keV soft X-ray flux is 
1.6 or 2.0$\,10^{-14}\ \tr{ergs}\,\unit{cm}{-2}\,\unit{s}{-1}$ (CPL or CTT).
An alternative model was also used to explain the origin of the
soft X-ray excess based on the idea that the absorbing matter could only
partially cover the X-ray emitting source (model {\tt pcfabs} in Xspec). A
covering factor of $0.995\pm 0.002$ could also explain the soft excess with a
good fit of $\chi^{2}/\mathrm{d.o.f.}=406/378$ for CTT model (411/378 for CPL).

\begin{table*}
\caption{Spectral analysis. The EPIC and ISGRI spectra were fitted together.
         Two models in {\tt Xspec} were selected to fit the data. Model CPL is
         {\tt constant*wabs(bbody+vphabs(cutoffpl+gaussian))}
         and model CTT is
         {\tt constant*wabs(bbody+vphabs(compTT+gaussian))}.
         The galactic absorption $N_{\tty{H}}^{\tty{gal}}$ was fixed. The errors
         are calculated at the 90\% confidence level.
	}             
\label{TabSpec}
\centering          
\begin{tabular}{l l c l} 
\hline\hline       
Models    & Parameters                &  Values                 & Unit \\
\hline                    
model CPL & $C_{\tty{ISGRI}}$         & 2.6                     &                         \\
	  & $N_{\tty{H}}^{\tty{gal}}$ & 1.5                     & $10^{22}\ \unit{cm}{-2}$\\
	  & $kT_{\mathrm{soft}}$      & 0.5			& keV (fixed) 		  \\
	  & $F_{1-2 \tty{keV}}$       & 1.6                     & $10^{-14}\ \tr{ergs}\,\unit{cm}{-2}\,\unit{s}{-1}$\\
	  & $N_{\tty{H}}$             & $12.8_{-1.3}^{+0.7}$    & $10^{22}\ \unit{cm}{-2}$\\
	  & $Z_{\mathrm{Fe}}$         & $1.5_{-0.2}^{+0.2}$     & $Z_{\odot}$             \\
	  & $\Gamma$                  & $0.02_{-0.10}^{+0.17}$  &                         \\
	  & $E_{\tty{c}}$             & $8.2_{-0.3}^{+0.4}$     & keV                     \\
	  & $E_{\tty{line}}$          & $6.400_{-0.013}^{+0.001}$ & keV \\
	  & $F_{\tty{line}}$          & $1.9_{-0.2}^{+0.2}$     & $10^{-4}\ \tr{ph}\,\unit{cm}{-2}\,\unit{s}{-1}$ \\
	  & unabs $F_{2-12 \tty{keV}}$& 1.5                     & $10^{-10}\ \tr{ergs}\,\unit{cm}{-2}\,\unit{s}{-1}$ \\
	  & unabs $F_{13-100 \tty{keV}}$ & 5.1                  & $10^{-10}\ \tr{ergs}\,\unit{cm}{-2}\,\unit{s}{-1}$ \\
	  & $\chi^{2}$/d.o.f.         & 401/376                 & \\
\hline
model CTT & $C_{\tty{ISGRI}}$         & 1.7                     &                         \\                  
	  & $N_{\tty{H}}^{\tty{gal}}$ & 1.5                     & $10^{22}\ \unit{cm}{-2}$\\
	  & $kT_{\mathrm{soft}}$      & 0.5			& keV (fixed) 		  \\
	  & $F_{1-2 \tty{keV}}$       & 2.0                     & $10^{-14}\ \tr{ergs}\,\unit{cm}{-2}\,\unit{s}{-1}$\\
	  & $N_{\tty{H}}$             & $15.3_{-1.0}^{+1.1}$    & $10^{22}\ \unit{cm}{-2}$\\                  
	  & $Z_{\mathrm{Fe}}$         & $1.4_{-0.3}^{+0.3}$     & Z$_{\odot}$             \\
	  & $kT_{0}$                  & 0.1		        & keV (fixed)             \\
	  & $kT_{\tty{e}}$            & $5.5_{-0.2}^{+0.2}$     & keV                     \\                  
	  & $\tau$                    & $7.8_{-0.5}^{+0.6}$     &                         \\                  
	  & $E_{\tty{line}}$          & $6.401_{-0.009}^{+0.005}$ & keV                   \\                  
	  & $F_{\tty{line}}$          & $2.1_{-0.3}^{+0.2}$     & $10^{-4}\ \tr{ph}\,\unit{cm}{-2}\,\unit{s}{-1}$ \\
	  & unabs $F_{2-12 \tty{keV}}$& 1.7                     & $10^{-10}\ \tr{ergs}\,\unit{cm}{-2}\,\unit{s}{-1}$ \\
	  & unabs $F_{13-100 \tty{keV}}$ & 4.8                  & $10^{-10}\ \tr{ergs}\,\unit{cm}{-2}\,\unit{s}{-1}$ \\
	  & $\chi^{2}$/d.o.f.         & 401/376                 & \\
\hline                  
\end{tabular}
\end{table*} 

For phase-resolved spectroscopy with EPIC/pn, the pulse profile was cut into three
phase intervals: 0.00--0.43 + 0.81--1.00 for the low intensity par;, 0.43--0.49 +
0.71--0.81 for the wings of the broad pulse; 0.49--0.71 for the pulse core.
The effective exposures for each phase resolved bin are 3383~s, 894~s and
1229~s, respectively. The phase-resolved spectra were extracted and the channel
bins were grouped to have at least 50 counts per channel for the low intensity
part and the pulse core, and 20 counts per channel for the pulse wings.
All the resulting spectra were fitted inside {\tt Xspec} using the CTT model again 
combined with the ISGRI spectrum to better constrain the spectral fit. No
significant variation of the spectral shape with the phase was found, 
excepting the change in the normalisation (see Table~\ref{TabSpecPhase}).
				    
\begin{table*}
\caption{Phase-resolved spectral analysis. We used the CTT model.
         All continuum and normalisation parameters were left free.
	 A few parameters were fixed as for the average spectrum
	 (see Table~\ref{TabSpec}): $N_{\tty{H}}^{\tty{gal}}=1.5\,10^{22}\ \unit{cm}{-2}$,
	 $kT_{\mathrm{soft}}=0.5\ \mathrm{keV}$, $kT_{0}=0.1\ \mathrm{keV}$.
	 Only pn data were used for the phase-resolved spectral analysis.
	}
\label{TabSpecPhase}
\centering          
\begin{tabular}{l c c c l} 
\hline\hline       
Parameters                & low intensity         & pulse wings           & pulse core            & Unit \\
\hline                    
$C_{\tty{ISGRI}}$         & 2.6                   & 1.4                   & 0.9                   & \\
$F_{1-2 \tty{keV}}$       & 1.4                   & 4.2                   & 2.8                   & $10^{-14}\ \tr{ergs}\,\unit{cm}{-2}\,\unit{s}{-1}$\\
$N_{\tty{H}}$	          & $15.2_{-2.2}^{+2.0}$  & $13.7_{-2.6}^{+2.9}$  & $15.1_{-1.8}^{+1.8}$  & $10^{22}\ \unit{cm}{-2}$ \\
$Z_{\mathrm{Fe}}$         & $1.2_{-0.5}^{+0.6}$   & $2.0_{-0.8}^{+1.2}$   & $1.6_{-0.4}^{+0.5}$   & Z$_{\odot}$ \\
$kT_{\tty{e}}$            & $5.5_{-0.2}^{+0.2}$   & $5.4_{-0.2}^{+0.2}$   & $5.5_{-0.2}^{+0.2}$   & keV \\		  
$\tau$  	          & $8.0_{-0.8}^{+1.0}$   & $8.7_{-1.2}^{+1.5}$   & $7.8_{-0.7}^{+0.8}$   & \\		  
$E_{\tty{line}}$          & $6.40_{-0.02}^{+0.01}$& $6.42_{-0.03}^{+0.04}$& $6.40_{-0.05}^{+0.04}$& keV \\
$F_{\tty{line}}$          & $2.2_{-0.3}^{+0.3}$   & $1.6_{-0.6}^{+0.6}$   & $1.4_{-0.6}^{+0.7}$   & $10^{-4}\ \tr{ph}\,\unit{cm}{-2}\,\unit{s}{-1}$ \\
unabs $F_{2-12 \tty{keV}}$& 1.2                   & 2.0                   & 3.4                   & $10^{-10}\ \tr{ergs}\,\unit{cm}{-2}\,\unit{s}{-1}$ \\
$\chi^{2}$/d.o.f.         & 145/160               & 208/217               & 180/197               & \\
\hline                  
\end{tabular}
\end{table*} 

Therefore, the three phase-resolved spectra were fitted with the Comptonization
model and all the continuum parameters were fixed, except its normalisation, the
column density, and the soft excess and the line normalisations. The results are
summarized in Table~\ref{TabPhaseSpec}. No significant
variation of the line flux nor of the absorbing column density was observed. The unabsorbed
2--10 keV flux changes according to the pulse phase. The variations of the
1--2 keV flux are not significant once the errors are considered.
	  
\begin{table*}
\caption{
         Phase-resolved spectroscopy with all the spectral modeling parameters
         of model CTT except normalisations equal to the values of the average
         spectrum (see Table~\ref{TabSpec}). Only pn data were used for the
         phase-resolved spectral analysis.
        }
\label{TabPhaseSpec}      
\centering          
\begin{tabular}{l c c c c c} 
\hline\hline       
phase & $N_{\tty{H}}$ & $F_{\tty{line}}$ & $F_{1-2\ \mathrm{keV}}$ & unabsorbed $F_{2-10\ \mathrm{keV}}$ & $\chi^{2}$/d.o.f \\
                 & $10^{22}\ \unit{cm}{-2}$ & $10^{-4}\ \tr{ph}\,\unit{cm}{-2}\,\unit{s}{-1}$ & $10^{-14}\ \tr{ergs}\,\unit{cm}{-2}\,\unit{s}{-1}$ & $10^{-10}\ \tr{ergs}\,\unit{cm}{-2}\,\unit{s}{-1}$ & \\
\hline                    
low intensity & 15.3$_{-0.7}^{+0.7}$ & 2.1$_{-0.3}^{+0.4}$ & 1.5$_{-1.2}^{+1.5}$ & 0.9 & 147/164 \\
pulse wings   & 15.8$_{-0.7}^{+1.3}$ & 1.5$_{-0.6}^{+0.7}$ & 6.8$_{-4.9}^{+0.4}$ & 1.4 & 214/221 \\
pulse core    & 15.5$_{-0.5}^{+0.5}$ & 1.6$_{-0.7}^{+0.5}$ & 2.8$_{-2.2}^{+2.2}$ & 2.4 & 181/201 \\
\hline                  
\end{tabular}
\end{table*}

\begin{figure}
  \centering
  \resizebox{\hsize}{!}{\includegraphics[angle=-90]{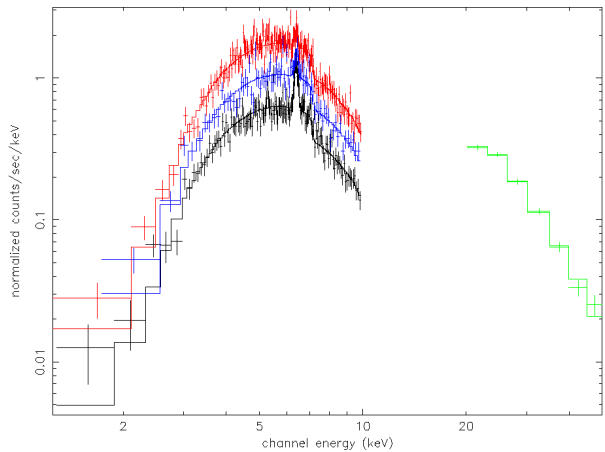}}
  \caption{
           EPIC/pn phase-resolved spectra. Only the pn spectra of the three phase
           intervals are displayed in the X-ray band: low intensity (bottom),
           pulse wings (middle) and pulse core (top). The ISGRI spectrum in the
           hard X-ray band is not phase resolved.
          }
  \label{FigPhaseSpec}
\end{figure}


\section{Discussion}\label{SecDisc}

\subsection{X-ray counterpart}

When \IGR\ was discovered in the $\gamma$-rays, it was first labelled as a new
source since no other known sources lay within the ISGRI error circle of
2$\arcmin$. The nearest source was \object{EXO~1722$-$363} at a distance of 12$\arcmin$.
The EXOSAT source was considered to be too far to be the X-ray counterpart of
\IGR. With the improved ISGRI and EPIC position, the source still lies 12$\arcmin$
away from the given EXOSAT position with an accuracy of 9$\arcmin$
\citep{Warwickal88}.

\citet{Tawaraal89} also gave a new position within a box
of $2\degr \times 4\degr$ that is 20$\arcmin$ away from the EPIC position.
They discovered a pulsation of 413.9~s. The pulse profile shows a single peak
that is independent of the energy and a high pulse fraction of 80\%. The
spectrum is heavily absorbed at low energy and shows a very hard shape plus
an emission line at 6.2$\pm$0.5 keV. Even if the position given by all these
studies is quite far from the accurate \IGR\ position, the informations given
by \citet{Tawaraal89} strongly confirm that the same source
is seen since the same features are seen in this study.

From archival data of RXTE/PCA taken between 1999 and 2003,
\citet{MarkwardtSwank03} reported an orbital period of 9.737$\pm$0.004 days,
consistent with the lower limit given by \citet{Takeuchial90}.
 This result with \emph{RXTE}\ was confirmed by \citet{Corbetal05}.
Their best orbital period is 9.741$\pm$0.004 d. They also find a pulse period of
413.88553$\pm$0.00001 s, equivalent to the previous studies and with the measurement presented here.

\subsection{Temporal modulations}

In this paper, periodic or episodic time variations were searched for on time
scales of seconds to days. First, the two years long ISGRI 20--60 keV
light curve indicates that the source is persistent. However, variations in the
$\gamma$-ray count rates larger than 50 were observed for this system. Four
flares were detected on a time scale of hours.

An orbital period of 9.72$\pm$0.09~days was derived that is consistent with
the one derived by \citet{Corbetal05}. The eclipse duration is a bit lower in
the ISGRI data than in the \emph{RXTE} data ($\sim$1.3~days and $\sim$1.7~days
respectively). Both orbital profile at 2--10 keV and 20--60 keV are consistent.
The progressive decrease of the intensity before entering the eclipse
between phase 0.7 and 0.97 can be explained by the changing hydrogenic column
density that increases from $\sim 10^{23}\ \unit{cm}{-2}$ to 
$\sim 10^{24}\ \unit{cm}{-2}$. This larger amount of matter along the line of
sight implies that a larger part of the flux is scattered.

Similar pulsations were found in EPIC/pn and ISGRI data: 414.8$\pm$0.5 s and
413.7$\pm$0.3 s, respectively. The EPIC and ISGRI pulsations are consistent with
the one detected by \citet{Tawaraal89} and \citet{Corbetal05}.
Therefore, the spin period did not change significantly in the last 17
years. \citet{Takeuchial90} also searched for pulse period changes and did not
find any significant spin variation in Ginga data. The neutron star seems to not
accrete kinetic momentum, which suggests accretion from a stellar wind.

With such spin pulsation and orbital period, the source is situated in the
underfilled Roche-Lobe supergiant region in the $P_{\mathrm{spin}}$ vs
$P_{\mathrm{orb}}$ Corbet diagram \citep{Corbet86}. Together with the fact that
the source is persistent and shows non-periodic flares with different
intensities, this strongly suggests that \IGR\ is a high mass
system fed by stellar wind with the primary star being a supergiant.

\subsection{Spectra}

The EPIC and ISGRI spectra that could be well fitted with a flat power law
($\Gamma\sim 0$) and an energy cutoff at $E_{\mathrm{c}}=8.2$ keV 
are typical of X-ray pulsars \citep{Whiteal95}. 
Three spectral features can give more clues about the physics of this
object: the apparent soft excess, the huge hydrogen column density and the
cold Fe K$\alpha$ line.

A soft excess is detected below 2~keV. Such soft X-ray excess has been
observed in other HMXB \citep{Hickoxal04}. It could originate in X-ray
scattering or partial ionization in the stellar wind \citep{Whiteal95}.
The spectrum at soft X-rays could also be explained by a partial
covering of the X-ray source by the absorbing matter.
In the latter case, pulsations at soft X-rays would be expected.
The 0.4--2 keV light curve was extracted and folded using the same ephemeris and
period used to fold the 0.4--10 keV light curve. Then, it was compared to a constant
model with $\chi^{2}/\mathrm{d.o.f.}\sim 57/50$ and the 0.4--10 keV pulse
profile with $\chi^{2}/\mathrm{d.o.f.}\sim 54/50$.
Both models are compatible with the 0.4--2~keV folded light curve.
However, the number of events gathered below 2~keV is too poor to decide if the
soft excess is pulsating or not.

The absorbing column density was estimated at 13--15$\,10^{22}$ atoms$\,\unit{cm}{-2}$
(model CPL-CTT) that is ten times larger than expected on the line of sight
$N_{\mathrm{H}}=1.5\,10^{22}\ \mathrm{atoms}\,\unit{cm}{-2}$. This high
absorption could be explained by the stellar wind expelled by the primary star
that surrounds the neutron star. During the \XMM\ observation, the column
density did not vary with the pulse. The fact that the amount of matter did not
change noticeably during the 10~ks \XMM\ observation could indicate that the
surrounding matter is stable on this time scale. However, other values of the
column density were reported at different epochs by \citet{Tawaraal89},
\citet{Takeuchial90} and \citet{Corbetal05}. The column density has been
observed to increase up to $10^{24}\ \mathrm{atoms}\,\unit{cm}{-2}$ in the past.
This could be related to the neutron star moving along the orbit
\citep{Tawaraal89,Corbetal05}. Therefore, the matter does not homogeneously
surround the binary system.
Instabilities in the stellar wind of the companion star could also be
responsible for this evolving absorption. However, in \citet{Corbetal05}, the
column density seems to be linked to the orbital phase where the highest values
correspond to the exit of the eclipse and the lowest ones when the source is in
front of the companion star.
The absorption is therefore intrinsic to the binary system.

The 6.4~keV Fe~K$\alpha$ line is detected. The Fe~K$\alpha$ line flux does
not vary with the pulse phase (see Table~\ref{TabPhaseSpec}). 
Therefore, either the matter responsible for the fluorescence is homogeneously
distributed around the source or the thickness of the shell emitting the
fluorescence is larger than $1.2\,10^{8}$~km around the accreting system.
The Fe~K$\alpha$ line energy is 6.401$_{-0.009}^{+0.005}$~keV (CTT model)
and corresponds to iron that is at most 12 times ionized \citep{House69}.
An upper limit of the ionization parameter can be estimated as $log(L/nR^{2})<1$,
where $L$ is the luminosity, $n$ the gas density, and $R$ is the distance from
the ionizing continuum source to the inner shell surface \citep{Kallmanal04}.
Since $nR\sim N_{\mathrm{H}}$, the distance of the fluorescence source from the
X-ray source is larger than $10^{7}$~km.
The values derived for the hydrogen column density, the equivalent width, and
the estimated over abundance of iron are compatible with a spherical
distribution of matter around the source \citep{Matt02}.

Considering a typical luminosity of an active accretion-powered pulsars of
1.2$\,10^{36}\,\mathrm{ergs}\,\unit{s}{-1}$ \citep{Bildstenal97}, the distance
of \IGR\ can be estimated as 7 kpc, close to the galactic centre.

\section{Conclusions}\label{SecConc}

\IGR\ is the hard X-ray counterpart of \object{EXO~1722$-$363}, and the most accurate
source position to date has been provided here. \IGR\ has been monitored by
\INTEGRAL\ during two years for a total exposure of 6.5~Ms and \XMM\ performed a
follow-up observation of 3~hours. The source is persistent with an average
20--60~keV flux of 6.4~mCrab.
Four flares lasting less than one day were detected by \INTEGRAL. The source's
count rate varies by a factor larger than 50 on such timescales. A pulsation
has been detected in both EPIC and IBIS/ISGRI data of 414.8$\pm$0.5~s and
413.7$\pm$0.3 s, respectively. There is no evidence of spin period variation.
An orbital period of 9.72$\pm$0.09 d is also found in IBIS/ISGRI data.
The spectral shape is typical for an accreting pulsar except that a huge
intrinsic absorption and a cold iron fluorescence line are detected. The
absorbing column density and cold iron line do not vary with the pulse period.
The absorbing fluorescent material is distributed around the neutron star in a
shell comparable in size with the orbital radius. With the accurate X-ray
position, we provide a likely infra-red counterpart within the X-ray error box:
\object{2MASS J17251139$-$3616575}. The observed features of the source suggest that it is
a wind-fed accreting pulsar. This object is a new member of the growing family
of heavily-obscured HMXB systems that have been recently discovered with
\INTEGRAL. The source is located $\sim$7~kpc away, near the Galactic Centre.

\begin{acknowledgements}

Based on observations obtained with the ESA science missions \INTEGRAL\ and
\XMM. The \INTEGRAL\ and \XMM\ instruments and data centres were directly funded
by ESA member states and the USA (NASA).
JAZH thanks J.Rodriguez for his useful comments and N.Produit for his help with
{\tt ii\_light}.

\end{acknowledgements}

\bibliographystyle{aa}
\bibliography{biblioIGR.bib}


\end{document}